# Degenerate Resistive Switching and Ultrahigh Density Storage in Resistive Memory

Andrew J. Lohn*, Patrick R. Mickel*, Conrad D. James, Matthew J. Marinella

*These authors contributed equally

Sandia National Laboratories, Albuquerque, New Mexico 87185, USA

Email: ajlohn@sandia.gov, prmicke@sandia.gov

Abstract:

We show that, in tantalum oxide resistive memories, activation power provides a multi-level variable for information storage that can be set and read separately from the resistance. These two state variables (resistance and activation power) can be precisely controlled in two steps: (1) the possible activation power states are selected by partially reducing resistance, then (2) a subsequent partial increase in resistance specifies the resistance state and the final activation power state. We show that these states can be precisely written and read electrically, making this approach potentially amenable for ultra-high density memories. We provide a theoretical explanation for information storage and retrieval from activation power and experimentally demonstrate information storage in a third dimension related to the change in activation power with resistance.

*ajlohn@sandia.gov

# 1. Introduction

Moore's law [1], the exponential increase in transistor density, has been the driving force for an immense increase in memory density over the past decades but it has become increasingly difficult to continue, leading to a broad search for new approaches [2-4]. Resistive memories (RRAM) [5,6] are among the leading candidates to replace transistors for memory applications, in part, due to their ability to store information within a range of resistances [7-9] instead of a digital ON or OFF. We show that by using partial SET followed by partial RESET operations, the information storage capacity of these devices can be increased by an order of magnitude or more. Previous observations of degeneracy in resistive switching are very few and have been limited to only a single level of degeneracy. Devices using two inert electrodes were able to create a vertical concentration gradient to provide a single level of degeneracy for storage. [10] And two turn-on voltages were observed at a single resistance value [11] but without establishing a mechanism for the degeneracy and without discussion of the potential for information storage.

Since RRAM requires Joule heating for thermal activation [12,13], the electrical power required to reach that activation threshold provides a separate state variable apart from resistance to store information. We experimentally demonstrate a simple electrical method for storing and extracting multiple activation power states at single resistance values. No changes are required to device design and, in principle, storing and extracting information in activation power states can be done simply. To compete with contemporary storage class memory however, new analog circuits will need to be developed to ensure, for example, speed and small footprint. In this paper we focus on explaining information storage using multiple dimensions (i.e. resistance and power) and on demonstrating that, in principle, it can be used to store a very large number of states per device.

For experimental demonstration, we used standard CMOS-compatible tantalum oxide metal-insulator-metal devices [14,15] with a reactive (tantalum) and an inert (titanium nitride) electrode. From a physical point of view the activation power and resistance are controlled by the radius and conductivity of the conducting filament [16]. The process is described in detail in three equivalent ways in Figure 1. The subfigures in the left column (Figure 2a,d,g) show the structure of the conducting filament, illustrating its radius and vacancy concentration (effectively conductivity). The subfigures in the middle column (Figure 2b,e,h) show the same changes on a current-voltage hysteresis loop. And the subfigures in the right column (Figure 2c,f,i) show the same current-voltage data mapped into power-resistance coordinates which are the two dimensions where information is stored. The electrical data can be directly converted between current-voltage and power-resistance coordinates via the equations $P = IV$ and $R = \frac{V}{I}$.

The physical process occurring within the device during switching is described in reference [16]. ON-switching (resistance decrease) proceeds in two stages as follows: First, applying a positive current (inert electrode grounded) decreases resistance by injecting oxygen vacancies from the reactive electrode into a localized conducting filament (Figure 2a-c). Next, once the injected vacancies reach a saturation concentration, further increases in positive current cause the radius of the filament to increase (Figure 2

2d-f) because injected vacancies cannot exist above saturation. To continue increasing the radius, more power must be supplied. In that way, a desired filament radius can be specified by choosing a power limit. In practice the power is typically limited by selecting a current limit for the electrical supply.

Once the radius is set, the conductivity can be set by applying the opposite polarity (negative on reactive electrode) to draw oxygen vacancies back into the reactive electrode, thereby increasing resistance (Figure 2g-i). To continue decreasing conductivity, more power must be supplied. In that way, the final conductivity can be specified by selecting a power limit. In practice the power is typically limited by setting a voltage limit. The ability to limit power by specifying a maximum voltage when resistance is increasing, and a maximum current when resistance is decreasing can be seen from the basic power relations and considering how the power supplied changes with changing resistance:

$$\text{Power} = V^2/R = I^2 R \qquad (1)$$

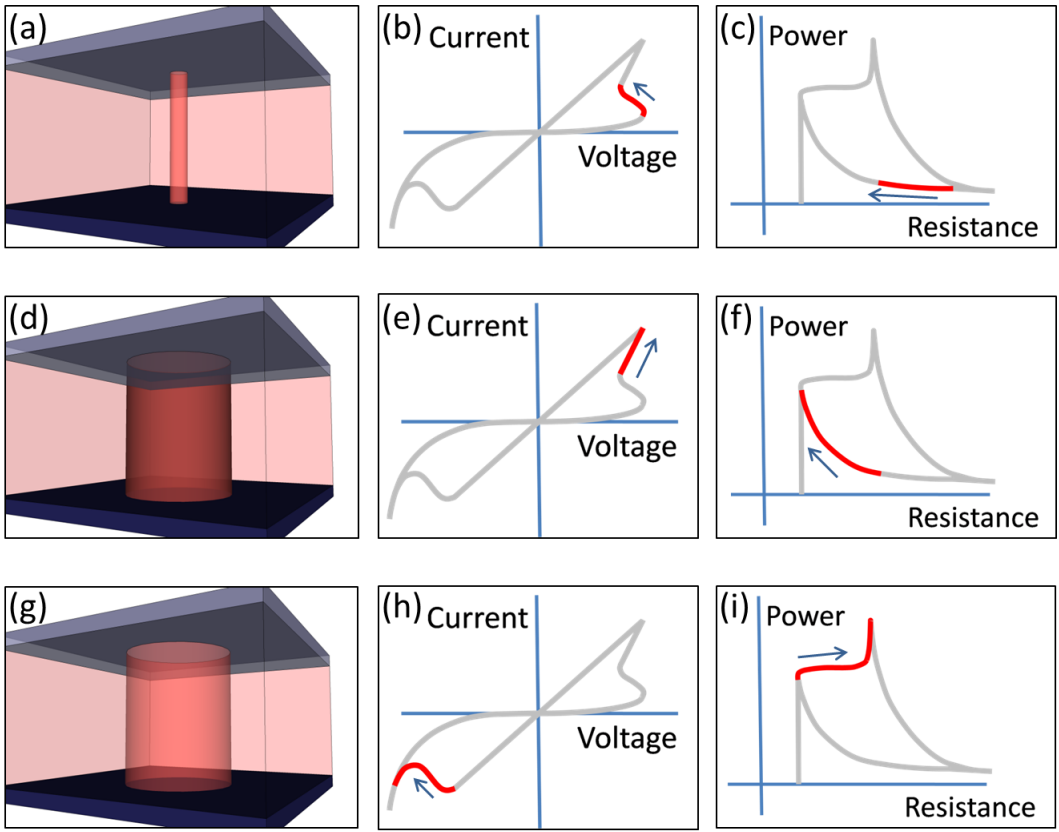

**Figure 1.** The radius and conductivity of a resistive filament can be tuned separately by progressing through the hysteresis loop which can be equivalently represented in either current-voltage (b,e,h) or power-resistance (c,f,i) coordinates. Starting from the high resistance state, positive polarity on the reactive electrode causes a filament to increase conductivity (a,b,c) until saturation. Further increasing power with positive polarity causes the radius to increase (d,e,f) and reversing polarity causes

conductivity to decrease (g,h,i). In this way various geometry-conductivity combinations can be achieved.

This two-step process (positive current then negative voltage) can be used to set a desired resistance in many different degenerate radius-conductivity combinations. For example, starting with a large current limit, then reversing polarity and using a large voltage limit will create a conducting filament with a large radius but small conductivity. The device will therefore have an intermediate resistance value. But that same resistance could alternatively be achieved by a conducting filament with a small radius and large conductivity. To create the latter filament one would use a small positive current to set a small radius then use a small negative voltage to only reduce the conductivity by a small amount. These two configurations have the same resistance but are fundamentally different.

It is difficult to directly measure the composition of the nanoscale filaments embedded in electronic devices but fortunately, the electrical behavior allows an immediate test for degenerate configurations. Using the physical description outlined above, a filament with a larger radius allows more heat to escape through the electrodes than a smaller filament of the same resistance. This simplified description predicts that large radius devices will require more power to initiate switching. Since the electrical resistance and electrical power are trivial to measure, they provide an electrical method to verify the existence of degenerate resistance states.

These power-related effects can be seen in a current-voltage hysteresis loop, but they are much more clearly observed if that same data is transformed into power and resistance coordinates. Experimental data taken from an example device is shown in Figure 2 in power-resistance coordinates. Prior to each of the curves shown, the device was set to a specified degenerate resistance value by applying a controlled positive current followed by a controlled negative voltage. The intersections of the dotted lines show the resistance and power at the end of the SET process. This is the location of the stored data in both power and resistance dimensions. It is clear from the figure that these set points are degenerate in both resistance and power since there are two distinct power configurations for each resistance and there are two distinct resistances for each power.

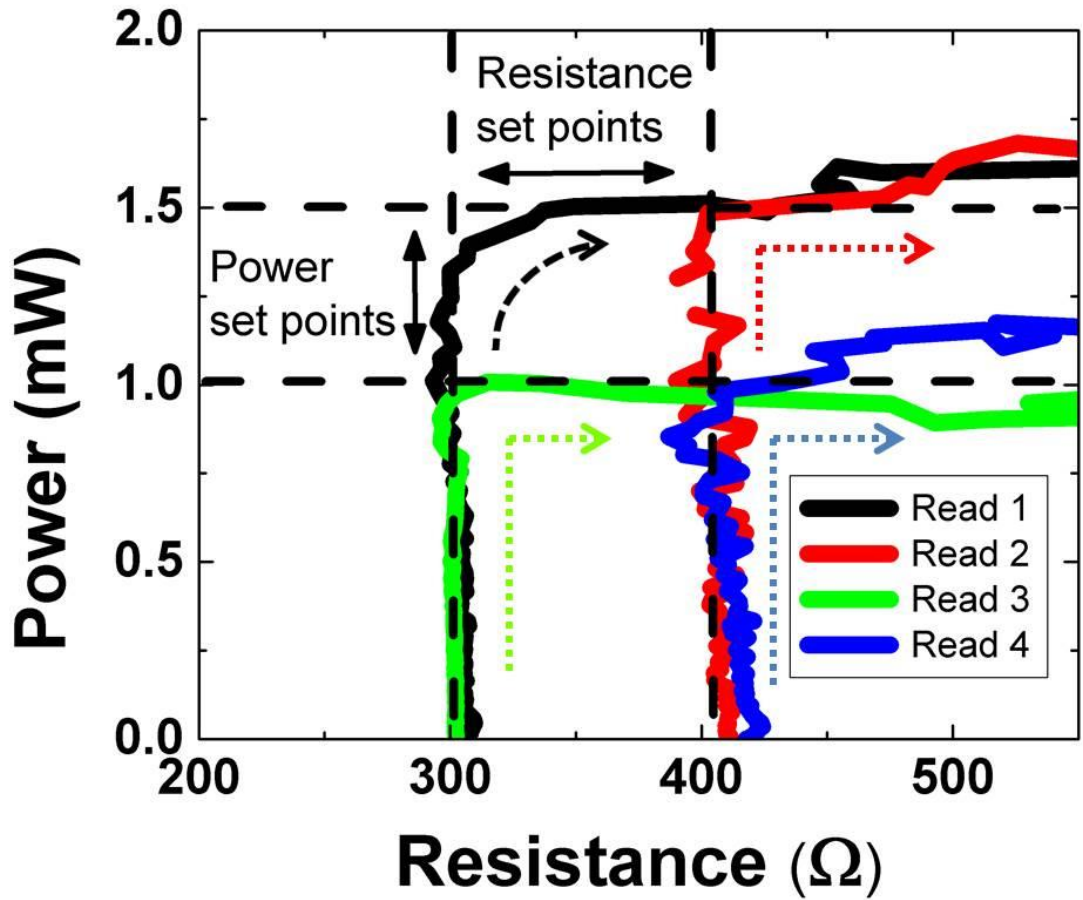

**Figure 2.** Information can be stored in both activation power and resistance by measuring resistance and observing the power required to change the resistance state. The SET points are indicated by the intersections of the dotted lines and the read points are the location of the kink in the solid curves (i.e. where sufficient power is delivered to activate resistance change).

In order to detect the configuration of the filament, the electrical power delivered to the device was progressively incremented while monitoring the resistance (moving upward along the solid curves). Once the activating power is reached, the resistance starts to change (the curve moves to the right). The kink in the curve is the READ location of the stored data. As seen in the figure, the locations of the kinks in the READ step are precise reproductions of the SET points (intersections of the dotted lines). The exception is the top left point where the READ curve activates at too low a power and departs gradually instead of abruptly like the others. This effect will be discussed later.

In Figure 2, only four distinct states are shown but both the resistance and power states can be set to a continuum of values. A schematic illustration of the range of possible values is shown in Figure 3a and an experimental demonstration is shown in Figure 3b. In the experimental demonstration, each point is set by pulses in the positive direction followed by pulses in the negative direction as described above. For the purpose of visual clarity in our experimental demonstration, we only show approximately 100 states but it is apparent from the figure that much larger state densities are possible. Even at this state density per dimension, this single device stores approximately 7 digital bits (i.e. $2^7$ states).

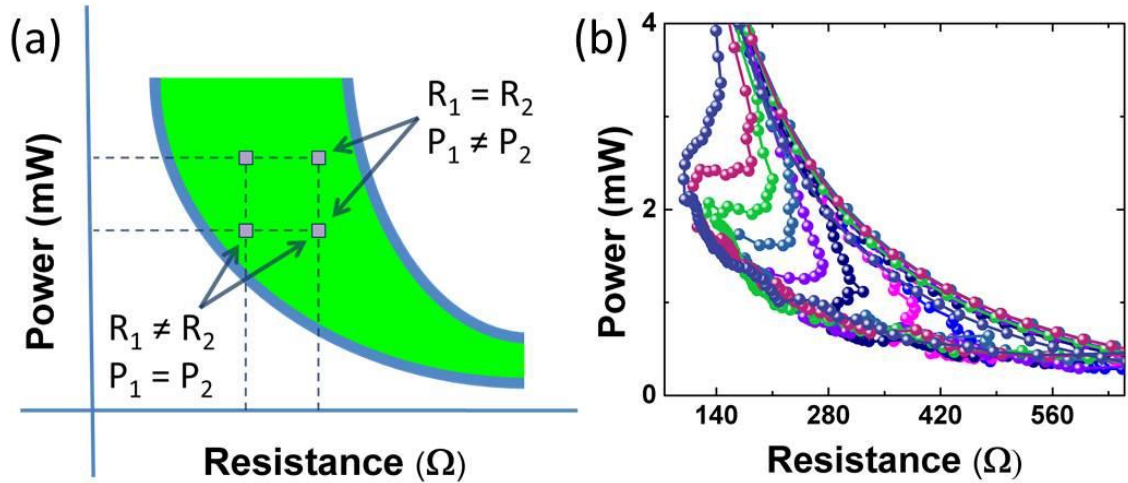

**Figure 3.** Information can be stored within any point in the green bounds in (a) illustrating the range of available states that can be achieved. A range of possible states is shown experimentally in (b) where approximately 100 states (7 digital bits) are shown in a single device.

As depicted in Figure 2, the READ points (location of the kinks) are very good reproductions of the SET points (intersections of dotted lines), indicating that high densities of information storage may be possible. We will now briefly discuss the discrepancy between the SET and READ points in the high-power low-resistance case (top left in the Figure 2). The early activation and more gradual resistance change is due to the shape of the thermal profile within the filament and can be described as such: for a cylindrical filament, the temperature is highest in the center so a small segment of the cylinder near the center will become activated prematurely, allowing for early onset of resistance change. Since only a fraction of the filament is thermally activated, the resistance changes gradually. The following equation (see Supporting Information for derivation) relating the SET and READ voltages required to reach thermal activation for a cylindrical filament with uniform conductivity provides qualitative insight:

$$V_{read} = V_{set} \sqrt{\frac{\frac{d_E}{k_E} - \frac{r_F^2}{4k_F d_o}}{\frac{d_E}{k_E} + \frac{r_F^2}{4k_F d_o}}} \qquad (2)$$

where $d_E$ and $d_o$ are the electrode and oxide thickness respectively, $k_E$ and $k_F$ are the electrode and filament thermal conductivity respectively, and $r_F$ is the filament radius.

Inspection of Eq 2 shows that if the following inequality holds, then the READ voltage reduces to the SET voltage.

$$\frac{d_E}{k_E} \gg \frac{r_F^2}{4k_F d_o} \qquad (3)$$

In our devices, this simplifying approximation is not justified for all $r_F$ but it may be possible to alter design to meet this criterion. The most convenient approach to do so involves increasing the thermal resistance of the electrode. It should be noted that the above relation and simplifying assumption were derived by modeling the thermal resistance with $k_E$ and $d_E$ as continuum parameters to define electrode thermal resistance but the vertical thermal resistance itself is the parameter of interest and may be dominated by interface and geometric spreading effects.

Even without device redesign, it is still possible to read and write at the same activation power over a wide range. The SET-READ discrepancy is due to a temperature profile that peaks in the center, but the temperature profile in the device spontaneously flattens during OFF-switching due to the following negative feedback process: In the first stages of OFF switching, the temperature is highest at the center so the electrical conductivity is reduced there first. Current density is then decreased in the central region, causing core temperature to drop and surrounding temperature to rise. This negative feedback process continues with increased switching, naturally producing a conductivity profile that gives a flattened temperature profile. For a device with a flattened temperature profile, the entire filament activates simultaneously, just as it does for a small radius filament (see Eq. 3).

Devices with low activation power points correspond to small radii which, when considering the inequality expressed in Eq 3, explains why the two low power points in Figure 2 activate abruptly at exactly the SET power. The two high power points correspond to large radii where Eq 3 is not justified, so the SET and READ powers are not necessarily identical but, as can be seen in Figure 2, in the high resistance case (top right) they are the same despite the large radius. That is because, during the previous two-step SET process, the high resistance case underwent the negative feedback process described above to give a flattened temperature profile. This is an example of using electrical control to flatten the temperature profile in order to achieve abrupt switching at exactly the SET voltage. Using this technique, over a reduced range of states, the SET and READ powers will be identical with abrupt transitions in the activation power-resistance space, even without the modification for increased vertical thermal resistance.

Interestingly, the gradual resistance change in the high-power low-resistance (top left) case in Figure 2 hints at the existence of a third dimension for information storage. The slope that is traversed through power-resistance space upon activation can be a tunable parameter apart from activation power and resistance themselves. For this third dimension, the SET process requires an additional step.

Once the device has been set to a two-dimensional state, applying a small positive polarity power that is only sufficient to activate the filament locally in the center will increase the conductivity of the filament in the center, leaving the surrounding regions unchanged. The conductivity of this inner filament can then be reduced with a small negative polarity power. By providing an oscillating signal of partial ON and OFF half cycles, it is possible to create a gradient of conductivities instead of a uniform cylindrical geometry as discussed previously.

Various conductivity profiles can be created to have the same resistance and activation power states but with non-cylindrical conductivity profiles. Once sufficient electrical power is supplied, their resistance will change but so will their conductivity profile. As their conductivity profile changes, so will their activation power. Changes in the activation power during switching cause these different configurations to follow different paths through resistance-power space that can be detected separately from the resistance and activation power themselves.

An experimental example is depicted schematically in Figure 4a and experimentally in Figure 4b, where a device was configured to have the same activation power and resistance but different paths through power-resistance space. Equivalently, the state variables can be thought of as conductivity, radius, and slope of the filament's conductivity profile. Since the switching begins at the same resistance and power point, the varied slopes represent a third dimension for information storage. It is easy to envision that controlling the curvature or other aspects of the path could present still additional dimensions that can all be SET and READ electrically.

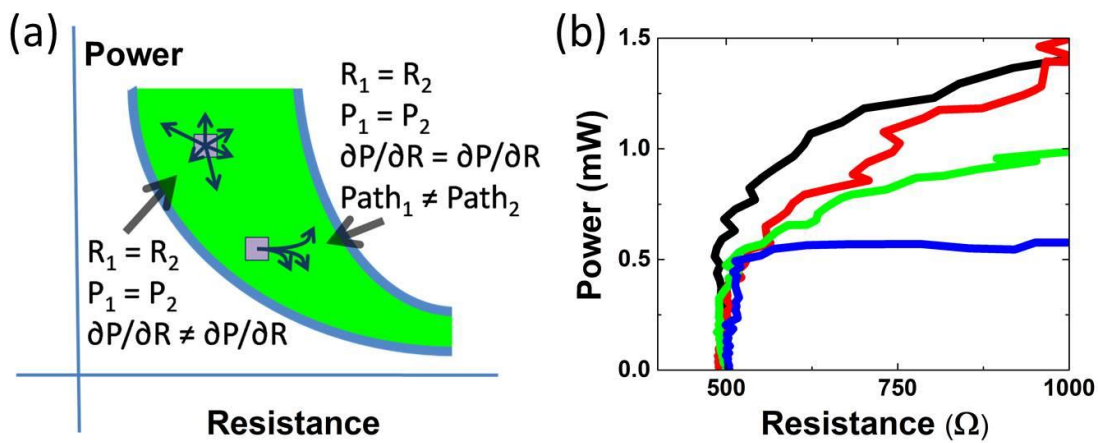

**Figure 4.** (a) Using the slope of the activation power-resistance curve once activation is achieved enables more dimensions for storage. These parameters can be controlled by intentionally altering the conductivity profile of the filament to be non-cylindrical. This effect is demonstrated experimentally (b) where the slope of the power-resistance path is separately written and read.

The ability to store and retrieve information in activation power as well as resistance was demonstrated experimentally and explained theoretically. The approach for writing and reading these variables requires only two steps to achieve the two dimensions. A positive polarity step is used to set the filament radius (effectively determining the power-state), and a negative polarity step is used to set the filament conductivity (effectively determining the resistance state). Storage in this multi-dimensional information space drastically increases the available state density of these devices without changes in device design.

## Acknowledgment

The authors would like to acknowledge James E. Stevens and the Sandia MESA Fab for device fabrication. This work was funded by Sandia's Laboratory Directed Research and Development program. Sandia National Laboratories is a multi-program laboratory managed and operated by Sandia Corporation, a wholly owned subsidiary of Lockheed Martin Corporation, for the U.S. Department of Energy's National Nuclear Security Administration under contract DE-AC04-94AL85000.

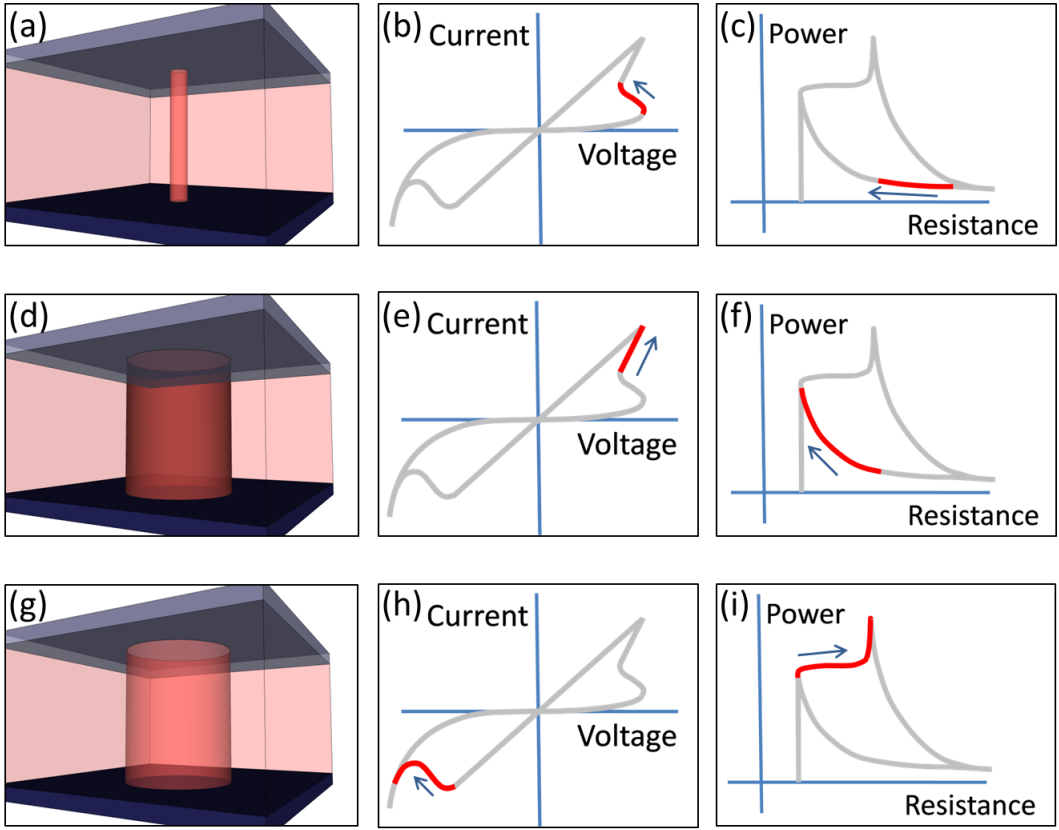

**Figure 1.** The radius and conductivity of a resistive filament can be tuned separately by progressing through the hysteresis loop which can be equivalently represented in either current-voltage (b,e,h) or power-resistance (c,f,i) coordinates. Starting from the high resistance state, positive polarity on the reactive electrode causes a filament to increase conductivity (a,b,c) until saturation. Further increasing power with positive polarity causes the radius to increase (d,e,f) and reversing polarity causes conductivity to decrease (g,h,i). In this way various geometry-conductivity combinations can be achieved.

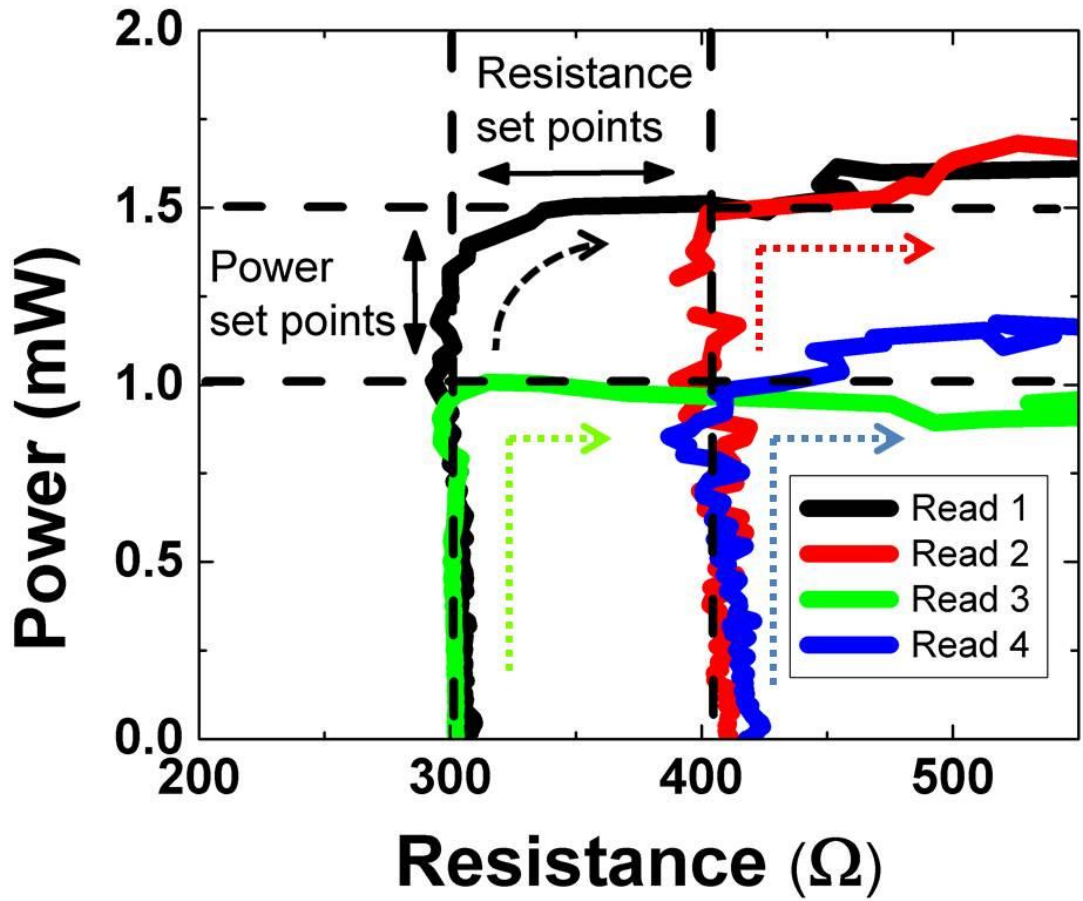

**Figure 2.** Information can be stored in both activation power and resistance by measuring resistance and observing the power required to change the resistance state. The SET points are indicated by the intersections of the dotted lines and the read points are the location of the kink in the solid curves (i.e. where sufficient power is delivered to activate resistance change).

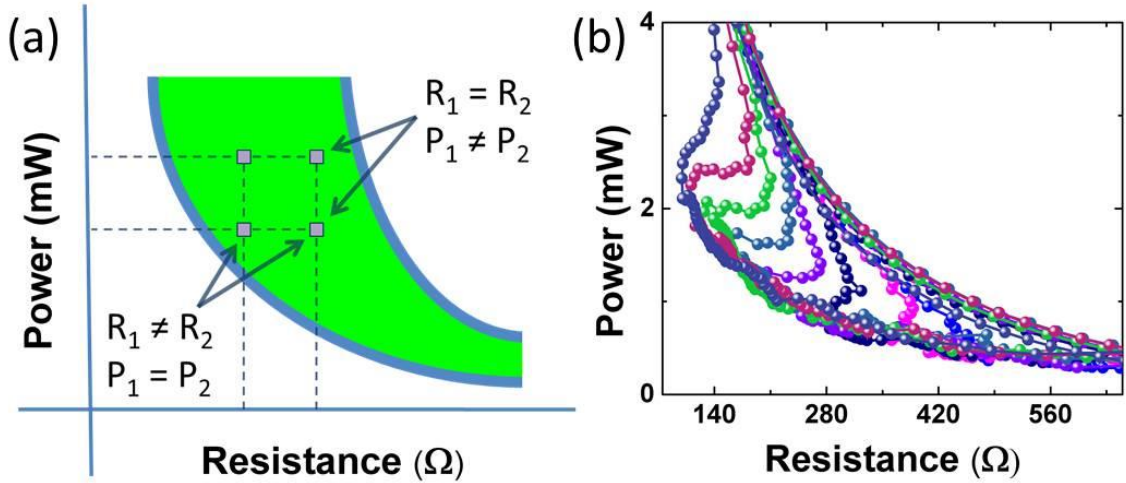

**Figure 3.** Information can be stored within any point in the green bounds in (a) illustrating the range of available states that can be achieved. A range of possible states is shown experimentally in (b) where approximately 100 states (7 digital bits) are shown in a single device.

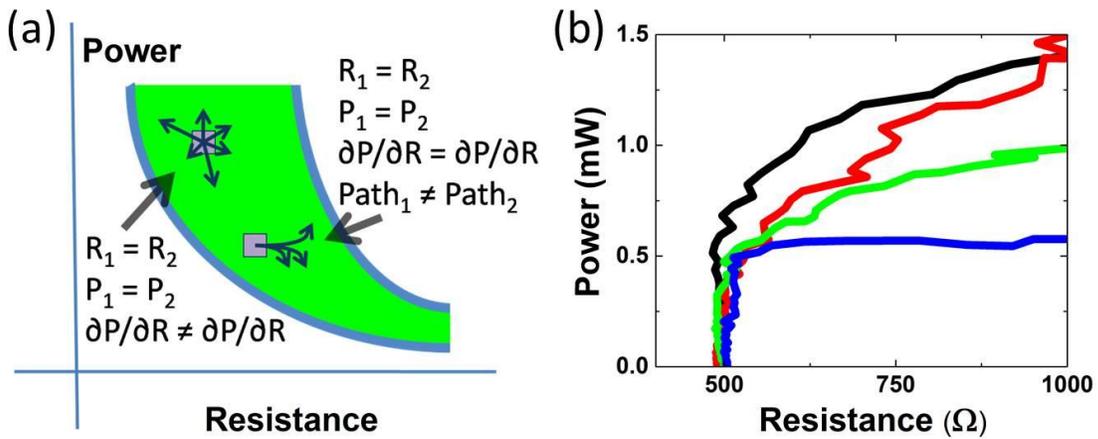

**Figure 4.** (a) Using the slope of the activation power-resistance curve once activation is achieved enables more dimensions for storage. These parameters can be controlled by intentionally altering the conductivity profile of the filament to be non-cylindrical. This effect is demonstrated experimentally (b) where the slope of the power-resistance path is separately written and read.

## Supporting Information

Derivation of Read Voltage:

During the SET process, the radius of the conducting filament increases once the oxygen vacancies reach a saturation concentration. Further increases occur once the outer edge of the filament reaches the activation temperature and vacancies there become mobile enough to enter. As a result, the conductivity within the filament is approximately uniform throughout, and the SET voltage is the voltage required to achieve the activation temperature at the edge of the filament, where new vacancies can enter. In a subsequent negative polarity step, vacancy motion is not limited by saturation so vacancies can leave from any portion of the filament that reaches the activation temperature. This will occur at the center of the filament first and at a lower voltage than is required to reach activation at the filament edge.

Equations given in Ref 15 approximate the temperature profile within a cylindrical filament with uniform conductivity. Equating the temperature at the edge (SET process and Eq S1) to the temperature at the center (read process and Eq S2) relates the SET and read voltages.

Temperature at the edge:

$$T_S = T_{RT} + \sigma V_{SET}^2 \frac{d_E}{2k_E d_O}\left[1 - \frac{k_E}{k_F}\frac{r_F^2}{4\,d_E d_O}\right] \quad (S1)$$

Temperature at every r within the filament:

$$T(r) = \frac{\sigma V_{read}^2 r_F^2}{4 d_O^2 k_F}\left(1 - \frac{r^2}{r_F^2}\right) + T_S \quad (S2a)$$

Temperature at r = 0:

$$T(r=0) = \frac{\sigma V_{read}^2 r_F^2}{4d_O^2 k_F} + T_{RT} + \sigma V_{read}^2 \frac{d_E}{2k_E d_O}\left[1 - \frac{k_E}{k_F}\frac{r_F^2}{4\,d_E d_O}\right] \quad (S2b)$$

Equating (1) and (2) and grouping terms gives:

$$V_{read}^2\left[\frac{\sigma r_F^2}{4d_O^2 k_F} + \frac{\sigma d_E}{2k_E d_O}\left[1 - \frac{k_E}{k_F}\frac{r_F^2}{4\,d_E d_O}\right]\right] = V_{SET}^2 \frac{\sigma d_E}{2k_E d_O}\left[1 - \frac{k_E}{k_F}\frac{r_F^2}{4\,d_E d_O}\right] \quad (S3)$$

$$V_{read} = V_{set}\sqrt{\frac{\frac{d_E}{k_E} - \frac{r_F^2}{4k_F d_o}}{\frac{d_E}{k_E} + \frac{r_F^2}{4k_F d_o}}} \quad (S4)$$

This equation relates the voltage required to increase resistance to the voltage required to decrease the resistance for a cylindrical filament with uniform conductivity. For a filament that has already been partially depleted, the two values will be more similar due to negative feedback during the initial stages of OFF switching. Additionally, the temperature profile within the filament that was used in the preceding derivation ignores heat loss through the electrodes, causing an overestimate of the peak temperature at the center. Therefore, the difference between SET and READ voltages in Eq S4 represents a maximum difference and should be thought of as limiting behavior that provides intuition in understanding the discrepancy between read and write.